\documentclass[twocolumn,pra,psfig,showpacs,superscriptaddress]{revtex4}

\usepackage{graphicx}
\usepackage{times}
\usepackage{color}

\usepackage{graphics}

\usepackage{amssymb}

\newcommand{\be}{\begin{equation}}
\newcommand{\ee}{\end{equation}}
\newcommand{\ben}{\begin{eqnarray}}
\newcommand{\een}{\end{eqnarray}}

\begin{document}

\title{Spontaneous symmetry restoration in a
field theory at finite chemical potential in a toroidal topology}

\author{C.A. Linhares{\footnote {linharescesar@gmail.com}}}
\affiliation{Instituto de F\'{\i}sica, Universidade do Estado do Rio
de Janeiro, 20559-900, Rio de Janeiro, RJ, Brazil}
\author{A. P. C. Malbouisson{\footnote {adolfo@cbpf.br}}}
\affiliation{Centro Brasileiro de Pesquisas F\'{\i}sicas/MCT,
22290-180, Rio de Janeiro, RJ, Brazil}
\author{J. M. C. Malbouisson{\footnote {jmalboui@ufba.br}}}
\affiliation{Instituto de F\'{\i}sica, Universidade Federal da
Bahia, 40210-340, Salvador, BA, Brazil}
\author{I. Roditi{\footnote {roditi@cbpf.br}}}
\affiliation{Centro Brasileiro de Pesquisas F\'{\i}sicas/MCT,
22290-180, Rio de Janeiro, RJ, Brazil}

\begin{abstract}
We consider the massive vector $N$-component
$(\lambda\varphi^{4})_{D}$ theory defined on a Euclidean space with
a toroidal topology. Using recently developed methods to perform a
compactification of a $d$-dimensional subspace at finite chemical
potential, we treat jointly the effects of temperature and spatial
boundaries, setting forth grounds for an analysis of spontaneous
symmetry restoration driven by temperature and spatial boundaries as
a function of the chemical potential. We restrict ourselves to $d=2$, which corresponds to the heated system confined between two parallel planes (separation $L$) in dimensions
$D=3$ and $D=4$. We present results, in the large-$N$ limit, which exhibit how finite size and chemical potential affect spontaneous symmetry restoration.
\end{abstract}


\maketitle

\section{Introduction}

We investigate spontaneous symmetry restoration induced by both
temperature and spatial boundaries. In particular, we are interested
in studying the influence of a finite chemical  potential in the
context of finite-size effects. We shall focus on the $N$-component
vector $\varphi ^{4}$ model within the large-$N$ approximation for
scalar fields. The symmetry restoration is carried out by starting
from the two-particle irreducible (2PI) formalism \cite{CJT,amelino}
in the Hartree--Fock approximation, which sums up all contributions
coming from daisy and superdaisy diagrams. We need then to implement
the 2PI formalism in a space with compactified dimensions at large
$N$, which allows to obtain nonperturbative corrections to the
coupling constant. For the sake of simplicity, we study the
particular case of finite temperature and one compactified spatial
coordinate. We can then, starting from the broken symmetry region,
show the behavior of the renormalized mass for different values of
the temperature, compactified spatial dimension and chemical
potential.

Field theories defined on spaces with some, or all, of its
dimensions compactified is of interest in several branches of
theoretical physics. For a Euclidean $D$-dimensional space, this
means that its topology is of the type $\Gamma
_{D}^{d}=(S^{1})^{d}\times \mathbb{R}^{D-d}$, or its counterpart
with the Minkowski signature, with $1\leq d\leq D$, $d$ being the
number of compactified dimensions. Each of these compactified
dimensions has the topology of a circle $S^{1}$. We refer to $\Gamma
_{D}^{d}$ as a {\it toroidal} topology. These theories are often
associated to extra spatial dimensions, as in particle physics,
where theories defined on toroidal spaces with extra spatial
dimensions are employed, for instance, as a  way to investigate the
electroweak transition and
baryogenesis~\cite{pani,pani1,pani2,pani3,pani6,pani7}. Also, recent
works involving the idea of extra spatial dimensions have been
performed in low-energy physics~\cite{(g-2)NPB,claudio}.

On the other side, in many cases, one is concerned with theories
defined on a space with a compactified subspace. A very important
development of this kind, which has its roots in the late fifties,
is the systematic approach to quantum field theory at finite
temperature, as proposed in~\cite {3mats1,3ume2,3ume3,3ume4}. An
analogous formalism can be constructed for compactified spatial
coordinates, in a $D$-dimensional Euclidean space. This is an idea
advanced for instance in~\cite{birrel1}. In this case, as it is
stressed in~\cite{kha1}, the compactification of spatial coordinates
can describe systems confined to limited regions of space, which are
interpreted as representing samples of material in the forms of
films, wires or grains. In the general case of quantum field
theories in toroidal topologies, that is, with compactification of
the time coordinate and of spatial dimensions, mathematical basis to
deal with this situation on general grounds are consolidated in
recent developments~\cite{AOP09,AOP11}. This provides a general
framework for the results from previous works where systems at
finite temperature and/or compactified spatial dimensions were
considered, both for bosonic systems, as for instance
in~\cite{Ademir,linhares1,JMP1,EPL12(1)} and for fermionic ones
\cite{3comp6,PRD12,AMM,AMM2,AMM3}.

The paper is organized as follows. In Section 2, symmetry
restoration is investigated along the lines of \cite{amelino}, which
establishes a formula for the renormalized mass. This formula is
then rewritten to take into account the compactification of the
imaginary time and one of the spatial dimensions, considering the
dependence with the chemical potential. We can then study the
behavior of the renormalized mass starting from the broken symmetry
region. In Section 3, corrections to the coupling constant, due to
compactification as well as the presence of a chemical potential are
discussed. In Section 4, we collect the results from the previous
sections and, specializing to spacetime dimensions $D=3$ and 4, we
fully determine the behavior of the critical temperature as a
function of the size of the system and the chemical potential. We
also present, in the same section, our conclusions and final
remarks.

\newpage

\section{Symmetry restoration in a toroidal space}

\label{sec2}

We consider the model described by the Lagrangian density
\begin{equation}
\mathcal{L} = \frac{1}{2}\partial _{\nu }\varphi _{a}\partial ^{\nu
}\varphi _{a}+\frac{1}{2}m^{2}\varphi _{a}\varphi
_{a}+\frac{u}{4!}\,(\varphi _{a}\varphi _{a})^{2}
\label{Lagrangiana}
\end{equation}%
in Euclidean $D$-dimensional spacetime, where $m$ and $u$ are
respectively the zero-temperature mass and the coupling constant in
absence of boundaries and at zero chemical potential.  Also, in Eq.~(\ref{Lagrangiana}), $\varphi_{a}$ 
are scalar fields, the sum over repeated indices is carried from $a=1$ to $N$ and $\nu=1$ to $D$.  We consider
the large-$N$ regime, such that $\lim_{N\rightarrow \infty
,u\rightarrow 0}(Nu)=\lambda$, with $\lambda $ fixed. To simplify
the notation,we drop out the $a$ indices, summation over them being
understood in field products.
In order to approach symmetry restoration for this model we first
follow the 2PI formalism developed in \cite{CJT,amelino}. In this
case, one finds a stationary condition for the effective action
in the Hartree--Fock approximation which translates itself into a
gap equation,
\begin{equation}
G^{-1}(x,y) = D^{-1}(x,y)+\frac{u}{2}\,G(x,x)\delta ^{4}(x-y),
\label{GAP}
\end{equation}
where the Fourier-transformed propagators, $D(k)$ and $G(k)$, are
given by
\begin{equation}
D(k) = \frac{1}{{k}^{2}+m^{2}+\frac{u}{2}\phi ^{2}}\,; \;\;\;\;G(k)
= \frac{1}{{k}^{2}+M^{2}}. \label{D}
\end{equation}
Here, $\phi = \left\langle 0\left\vert \varphi \right\vert
0\right\rangle $ is the expectation value of the quantum field
$\varphi $ and $M$ is a momentum-independent effective mass.

In the 2PI formalism, the gap equation corresponds to the stationary
condition and as such the effective mass depends on $\phi $ and
conveys all daisy and superdaisy graphs contributing to $G(k)$
\cite{CJT,amelino}. Nevertheless, in order to investigate symmetry
restoration, we can take instead a particular constant value $M$ in
the spontaneously broken phase. Renormalization of the mass $m$ and
of the coupling constant $u$ can be performed along lines similar to
those in~\cite{amelino,coleman}, leading to the equation
\begin{equation}
{M^{2} = -m_{R}^{2} + \frac{u_{R}}{2}\phi ^{2} +
\frac{u_{R}}{2} \, G(M)}, \label{M}
\end{equation}
where $m_{R}^{2}$ and $u_{R}$ are respectively the squared
renormalized mass and the renormalized coupling constant, both at
zero temperature and zero chemical potential, in the absence of
boundaries, {and $G(M)$ is the finite part of the integral $G(x,x) =
(1/(2\pi)^{D})\int d^{D}{k}\,G(k)$, which will be calculated using
dimensional and zeta-function regularization techniques}. Notice
also the minus sign of the $m_{R}^{2}$ term, which is the same
choice made in~\cite{amelino,coleman} to ensure spontaneous symmetry
breaking. Therefore, Eq.~(\ref{M}) will be our starting point. It
gives the value of the renormalized mass at the broken-symmetry
phase and can be rewritten as
\begin{equation}
{\bar{m}}^{2}(\phi ) = -M^{2}+\frac{u_{R}}{2}\int
\frac{d^{D}\,{k}}{(2\pi )^{D} }\frac{1}{{k}^{2}+M^{2}},
\label{lagrangiana1}
\end{equation}
where the \textit{effective renormalized mass} ${\bar{m}}^{2}(\phi
)=-m_{R}^{2}+(u_{R}/2)\phi ^{2}$ has been introduced. In the sequel
we will obtain the generalization of the above equation in such a
way as to include the toroidal topology as well as the chemical
potential, but first we notice that restoration of the symmetry will
occur at the set of points in the toroidal space where ${\bar{m}}
^{2}$ is null.

We now proceed to generalize Eq.~(\ref{lagrangiana1}) to a theory
defined on a space with a toroidal topology. In the general case,
the system is in thermal equilibrium with a reservoir at temperature
$\beta ^{-1}$ and confined to a $(d-1)$-dimensional spatial
rectangular box of sides $L_{j}$, $ j=2,3,...,d$. We use Cartesian
coordinates $\mathbf{r}=(x_{1},...,x_{d}, \mathbf{z})$, where
$\mathbf{z}$ is a $(D-d)$-dimensional vector, with corresponding
momentum $\mathbf{k}=(k_{1},...,k_{d},\mathbf{q})$, $\mathbf{q} $
being a $(D-d)$-dimensional vector in momentum space. Then the
Feynman rules should be modified according
to~\cite{kha1,AOP09,AOP11}
\begin{eqnarray}
&&\int \frac{dk_{\tau }}{2\pi }\rightarrow \frac{1}{\beta
}\sum_{n_{\tau }=-\infty }^{\infty },\qquad k_{\tau }\rightarrow
\frac{2n_{\tau }\pi }{
\beta }-i\mu ,  \nonumber \\
&&\int \frac{dk_{i}}{2\pi }\rightarrow
\frac{1}{L_{i}}\sum_{n_{i}=-\infty }^{+\infty },\qquad
k_{i}\rightarrow \frac{2n_{i}\pi }{L_{i}} ,\;\;i=2,3,...,d,\nonumber \\
\label{Matsubara}
\end{eqnarray}
where $\tau $ corresponds to imaginary time and $\mu$ is the
chemical potential. We consider the simpler situation of the system
at temperature $\beta ^{-1}$ and one compactified spatial coordinate
($x_2$) with a compactification length $L_{2}\equiv L$. In this
case, using Eq.~(\ref{Matsubara}), we can perform a suitable
generalization of the procedure in~\cite{amelino}, to take into
account finite-size, thermal and boundary effects in
Eq.~(\ref{lagrangiana1}). The integral over the $D$-dimensional
momentum in Eq.~(\ref{lagrangiana1}) becomes a double sum over
$n_{\tau}$ and $n_2\equiv n_x$ together with a ($D-2$)-dimensional
momentum integral.  Then, following steps similar as
in~\cite{Ademir} and using dimensional regularization to perform the
integral~\cite{Ramond}, the renormalized ($\beta,L,\mu$)-dependent
mass in the large-$N$ limit can be written in the form
\begin{widetext}
\begin{eqnarray}
m^{2}(T,L,\mu ) & = & - M^{2} + \frac{u_{R} M^{D-2}}{2} \frac{\pi
^{(D-2)/2}}{4\pi^{2}}\frac{\Gamma \left( s-\frac{D-2}{2}\right)
}{\Gamma\left( s\right) } \sqrt{a_{\tau } a_{x}} \nonumber \\
&&\left. \times \sum_{n_{\tau },n_{x}=-\infty }^{\infty }\left[
a_{\tau }\left( n_{\tau } - \frac{i\beta }{2\pi }\mu \right)
^{2}+a_{x}n_{x}^{2}+c^{2} \right] ^{(D-2)/2-s}\right\vert _{s=1},
\label{potefet5}
\end{eqnarray}
\end{widetext}
where we have changed variables in the integral, {$k_{i}/2\pi M
\rightarrow q_{i}$, and introduced the dimensionless quantities
$a_{\tau }=(M\beta)^{-2}$, $a_{x}=(M L)^{-2}$ and $c=(2\pi)^{-1}$}.
The double sum in Eq.~(\ref{potefet5}) is recognized as one of the
inhomogeneous Epstein--Hurwitz zeta functions,
$Z_{2}^{c^{2}}(s-\frac{D-2}{2} ;a_{\tau}.a_x;b_{\tau},b_{x})$, which
has an analytical extension to the whole complex
$s$-plane~\cite{Kirsten,Elizalde}; in general for $j=1,\,2$,
\begin{widetext}
\begin{eqnarray}
Z_{2}^{c^{2}}(\nu ;\{a_{j}\};\{b_{j}\}) =  \frac{\pi |c|^{2 -
2\nu}\,\Gamma(\nu - 1)}{\Gamma(\nu) \sqrt{a_1 a_2}}  + \frac{4
\pi^{\nu} |c|^{1 - \nu}}{\Gamma(\nu) \sqrt{a_1 a_2}} \left[
\sum_{j=1}^{2} \sum_{n_j = 1}^{\infty} \cos(2\pi n_j b_j) \left(
\frac{n_j}{\sqrt{a_j}} \right)^{\nu - 1} K_{\nu - 1}
\left(\frac{2\pi c\, n_j}{\sqrt{a_j}} \right) \right.  \nonumber \\
\left. + \, 2 \sum_{n_1,n_2=-\infty}^{\infty} \cos(2\pi n_1 b_1)
\cos(2\pi n_2 b_2)  \left( \sqrt{\frac{n_1^2}{a_1} +
\frac{n_2^2}{a_2}} \right)^{\nu - 1}  K_{\nu -1}\left( 2 \pi c\,
\sqrt{\frac{n_1^2}{a_1} + \frac{n_2^2}{a_2}} \right) \right] .
\label{Z2}
\end{eqnarray}
\end{widetext}
For us, $a_1=a_{\tau}$, $a_2=a_x$, $b_1=b_{\tau}=i\beta \mu /2\pi $,
$b_2=b_{x}=0$, $c=1/2\pi$ and $\nu=s-(D-2)/2$. Replacing
Eq.~(\ref{Z2}) into Eq.~(\ref{potefet5}), the thermal and boundary
corrected mass is obtained in terms of the original variables,
$\beta,L,\mu$ and of the fixed renormalized zero-temperature
coupling constant in absence of boundaries,
$$\lambda_R=\lim_{N\rightarrow\infty,u_R\rightarrow 0}(N\,u_R).$$
However, the first term in Eq.~(\ref{Z2}) implies that the first
term in the corrected mass is proportional to $\Gamma(1-D/2)$, which
is divergent for even dimensions $D\geq 2$~\cite{Ademir}. This term
is suppressed by a minimal subtraction, leading to a finite
effective renormalized mass; for the sake of uniformity, this polar
term is also subtracted for odd dimensions, where no singularity
exists, corresponding to a finite renormalization.

In dimension $D$, the renormalized zero-temperature  coupling
constant in absence of boundaries $\lambda_R$ has dimension of
$mass^{4-D}$; accordingly, we introduce a dimensionless coupling
constant $\lambda_R^{\prime}=\lambda_R\,M^{D-4}$. Also, we define
the dimensionless reduced temperature $t$, reduced chemical
potential $\omega$,  and the reduced inverse length of the system
$\chi$, in such a way that we have for any dimension $D$, the set of
dimensionless parameters defined  by
\begin{equation}
\lambda_R^{\prime}=\lambda_R\,M^{D-4}\,,\;\;\;t =
T/M\,,\,\,\chi=L^{-1}/M\,,\;\;\;\omega=\mu/M.
\label{reduzidos}
\end{equation}
We then can obtain, after subtraction of the polar term, which does
not depend on $\beta$, $L$ and $\mu$, the corrected mass,
$\bar{m}^{2}(D,\beta,L,\mu )$. This implies that the condition for
symmetry restoration, $\bar{m}^{2}(D,\beta,L,\mu )=0$, can be
written in terms of the above dimensionless parameters, replacing
$\lambda^{\prime}_R$ by the corrected coupling constant
$\lambda^{\prime}_R(D,\beta,L,\mu)$ (this is  precisely defined in
the next section, Eq.~(\ref{lambdaR1})) , in such a way that the
critical equation reads
\begin{widetext}
\begin{eqnarray}
 - 1 +\frac {\lambda_R^{\prime}(D,\beta,L,\mu)}{(2 \pi)^{D/2}}
\left[  \sum_{n=1}^{\infty }\cosh \left(\frac{ \omega  n}{t}\right)
\left(\frac{t}{n}\right)^{\frac{D}{2}-1}
K_{\frac{D}{2}-1}\left(\frac{n}{t}\right)  + \sum_{l=1}^{\infty }
\left(\frac{\chi}{l}\right)^{\frac{D}{2}-1}K_{\frac{D}{2}-1}
\left(\frac{l}{\chi}\right) \right. \nonumber \\
+ \left. 2 \sum_{n,l=1}^{\infty }\cosh \left( \frac{\omega
n}{t}\right) \left(\frac{1}{\sqrt{\frac{n^2}{t^2} +
\frac{l^2}{\chi^2}}}\right)^{\frac{D}{2}-1} \, K_{\frac{D}{2}-1}
\left( \sqrt{\frac{n^2}{t^2}+\frac{l^2}{\chi^2}}\right) \right] =0.
\label{massa14}
\end{eqnarray}
\end{widetext}

\section{Corrections to the coupling constant}

In this section, we follow the reasoning made in \cite{JMP1},
appropriately modified to incorporate the effects from the chemical
potential. We consider the zero-external-momenta four-point
function, which is the basic object for our definition of the
renormalized coupling constant. At leading order in $1/N$ it is
given by the sum of all chains of one-loop diagrams with four
external legs, which leads to the expression (we consistently define
$u_R^{\prime}=u_R\,M^{D-4}$)~{\cite{ZJ}}
\begin{equation}
\Gamma _{D}^{(4)}(0,\beta ,L,\mu) = \frac{u^{\prime}_{R}}{1+
Nu^{\prime}_{R}\Pi (D,\beta ,L,\mu )} , \label{4-point1}
\end{equation}
where the dimensionless one-loop diagram is given by
\begin{equation}
\Pi (D,\beta ,L,\mu ) = \left. \frac{\sqrt{a_{\tau } a_{x} }}{16
\pi^4} \sum_{n_{\tau },n_{x}=-\infty }^{\infty
}{\cal{I}}_{n_{\tau}n_x} (s) \right|_{s=2} , \label{lagrangiana3}
\end{equation}
where
\begin{equation}
{\cal{I}}_{n_{\tau}n_x}(s)=\int
\frac{d^{D-2}q}{ \left[ \mathbf{q}^{2}+a_{\tau }\left( n_{\tau
}-\frac{i\beta }{2\pi }\mu \right) ^{2}+a_{x}n_{x}^{2} +
c^{2}\right] ^{s}}. \nonumber \\
\end{equation}
Then, proceeding along lines similar to those leading to
Eq.~(\ref{massa14}), we  write $\Pi (D,\beta ,L,\mu)$ in the form
$$\Pi (D,\beta ,L,\mu) = H(D) + \frac{1}{(2 \pi)^{D/2}}R(D,\beta ,L,\mu ),$$
where $R(D,\beta
,L,\mu )$ is given by
\begin{widetext}
\begin{eqnarray}
R(D,\beta ,L,\mu )  =
 \sum_{n=1}^{\infty }\cosh \left(\frac{\omega  n}{t}\right)
\left(\frac{t}{n}\right)^{\frac{D}{2}-2}K_{\frac{D}{2}-2}
\left(\frac{n}{t}\right) + \sum_{l=1}^{\infty }
\left(\frac{\chi}{l}\right)^{\frac{D}{2}-2}   K_{\frac{D}2{}-2}
\left(\frac{l}{\chi}\right)  \nonumber \\
 + 2 \sum_{n,l=1}^{\infty }\cosh
\left(\frac{\omega n}{t}\right)\left(\frac{1}{\sqrt{\frac{n^2}{t^2}
+\frac{l^2}{\chi^2}}}\right)^{\frac{D}{2}-2} K_{\frac{D}{2}-2}\left(
\sqrt{\frac{n^2}{t^2}+\frac{l^2}{\chi^2}}\right)
\nonumber \\
\label{GG4}
\end{eqnarray}
\end{widetext}
and $H(D)\propto \Gamma \left( 2-\frac{D}{2}\right)$ is a polar
parcel coming from the first term in the analytic extension of the
zeta function in Eq.~(\ref{Z2}). Notice that from general properties
of Bessel functions,  the above equations are meaningful for a
reduced chemical potential satisfying the condition $0\leq\omega<1$.
This is the allowed range of $\omega$ in Eqs.~(\ref{massa14}) and
(\ref{GG4}). We see that for even dimensions $D\geq 4$, $H(D)$ is
divergent, due to the pole of the gamma function. Accordingly, this
term must be subtracted to give the renormalized single bubble
function $\Pi_{R}(D,\beta ,L,\mu )$. We get, simply, $\Pi
_{R}(D,\beta ,L,\mu ) = [1/(2 \pi)^{D/2})]R(D,\beta ,L,\mu )$. As
mentioned before, the term $H(D)$ is also subtracted in the case of
odd dimensions $D$, where no poles are present, corresponding to a
finite renormalization. Using properties of the Bessel functions, we
see that, for any dimension $D$ and finite values of the chemical
potential $\mu $,  $R(D,\beta ,L,\mu )$ satisfies the conditions
$\lim_{\beta ,L\rightarrow \infty }R(D,\beta ,L,\mu )=0$,
$\lim_{\beta ,L\rightarrow 0}R(D,\beta ,L,\mu )\rightarrow \infty $,
and $R(D,\beta ,L,\mu )>0$ for finite $\mu $ and for any values of
$D$, $\beta $ and $L$. Under these conditions, we
 define the dimensionless $\beta$-, $L$- and $\mu$-dependent
renormalized coupling constant $\lambda^{\prime} _{R}(D,\beta ,L,\mu
)$ at the leading order in $1/N$ as $\lambda^{\prime} _{R}(D,\beta
,L,\mu )\equiv N\Gamma _{D}^{(4)}(0,\beta ,L,\mu )$, which, from
Eq.~(\ref{4-point1}), after subtraction of the polar term $H(D)$ in
$\Pi (D,\beta ,L,\mu )$, leads to
\begin{equation}
\lambda^{\prime}_{R}(D,\beta ,L,\mu ) = \frac{\lambda
^{\prime}_{R}}{1+\lambda^{\prime}_{R} [1/(2 \pi)^{D/2})]{R}(D,\beta
,L,\mu )}. \label{lambdaR1}
\end{equation}
In the next section we will investigate the restoration of symmetry,
taking into account thermal, boundary and  finite chemical potential
corrections to the coupling constant as presented above.

\begin{figure}[th]
\begin{center}
\includegraphics[{height=11.0cm,width=7.0cm}]{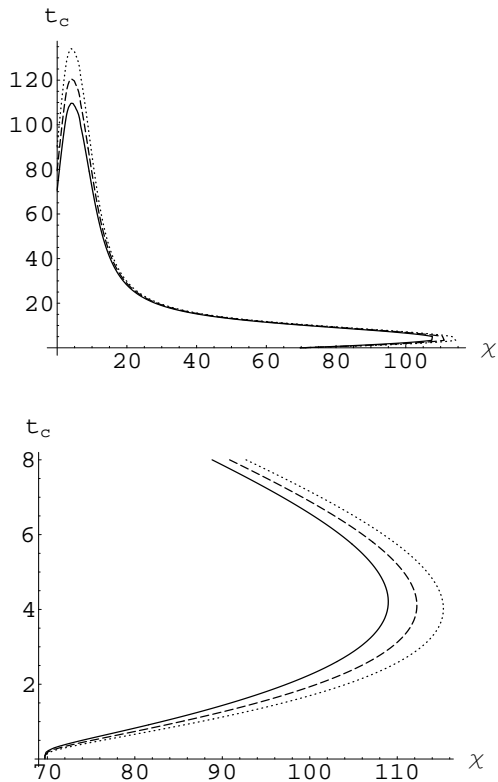}
\end{center}
\caption{{Reduced critical temperature as a function of the reduced
inverse size of the system for dimension $D=3$ (upper plot). We fix
$\protect\lambda^{\prime}_R=0.5$ and take the chemical potential
values $\protect \omega =0.1$ (full line), $0.3$ (dashed line) and
$0.4$ (dotted line). The symmetry-breaking regions are in the
``inner" side of each curve. The lower plot is a ``zoom" enhancing
the region of the characteristic size of the system, corresponding
to $\chi \approx 69.96$.}} \label{figPhi41}
\end{figure}

\begin{figure}[htp]
\begin{center}
\includegraphics[{height=11.0cm,width=7.0cm}]{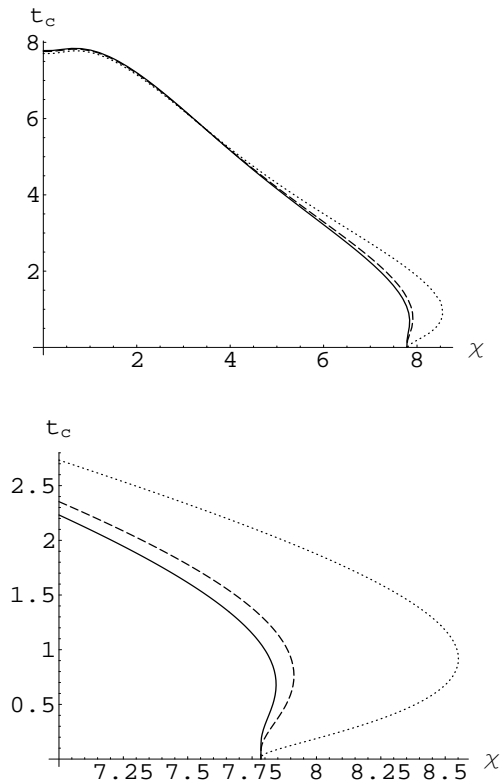}
\end{center}
\caption{{Reduced critical temperature as a function of the reduced
inverse size of the system  for dimension $D=4$ (upper plot). We fix
the (dimensionless) coupling constant $\protect\lambda_R =0.5$ and
take the chemical potential values $\protect \omega =0.0$ (full
line), $\omega=0.5$ (dashed line) and $\omega=0.9 $ (dotted line).
The symmetry-breaking regions are on the ``inner" side of each
curve. The lower plot shows in detail the critical temperature in
the region around the characteristic size of the system.} }
\label{figD4x}
\end{figure}

\begin{figure}[ht]
\begin{center}
\includegraphics[{height=5.5cm,width=8.0cm}]{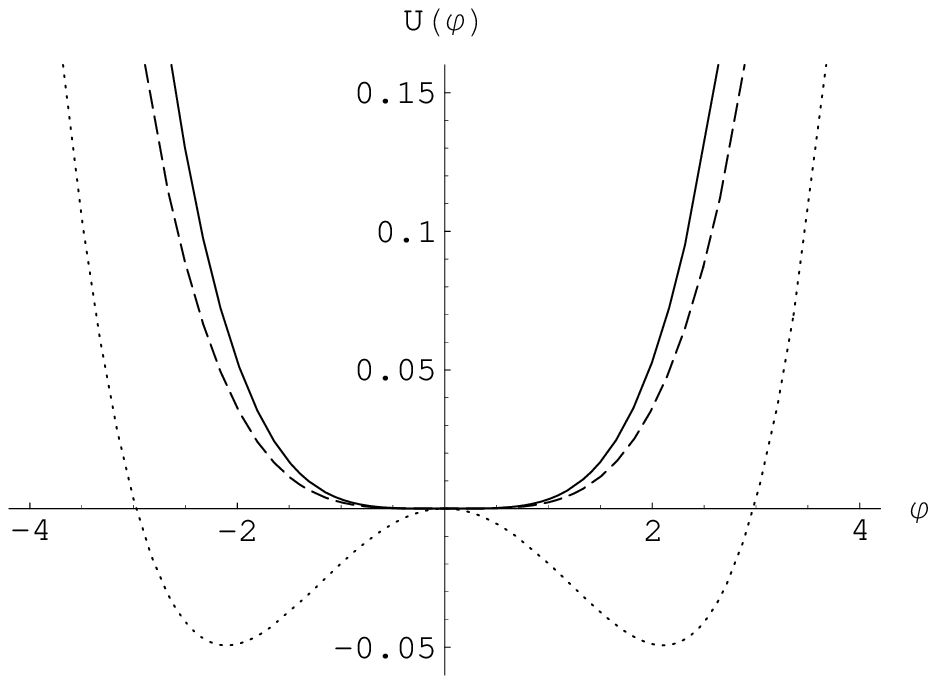}
\end{center}
\caption{{Curves for the effective potential
$U=\frac{1}{2}m^{2}\varphi^{2}+\frac{1}{24}\lambda
^{\prime}_{R}\varphi ^{4}$, for  fixed values of the reduced
chemical potential, $\omega=0.30$ and of the reduced inverse size of
the system, $\chi=100$ for $D=3$; $\varphi$ and $U$ are measured in
units of $M^{\frac{1}{2}}$ and $M^3$, respectively. We take three
values of the reduced temperature, two of them being critical
temperatures, $t_c^{(1)}=1.93$ (full line) and $t_c^{(2)}=6.89$
(dashed line), and an intermediate one, $t=4.80$ (dotted line), in
the symmetry-broken region. }} \label{figPhi44}
\end{figure}

\section{Boundary and chemical potential effects on the symmetry
restoration: comments and concluding remarks}

{In the general situation, Eq.~(\ref{massa14}) does not allow an
algebraic solution. For numerical evaluations, we fix the value
$\lambda^{\prime} _{R}=0.50$ and take several values of the
dimensionless parameters $t$, $\chi$ and $\omega$. We present and
comment on the results conccurently for dimensions $D=3$ and $D=4$,
since the behavior of the system is qualitatively the same, apart
from numerical details, in both cases.

In Figs.~\ref{figPhi41} and \ref{figD4x}  we exhibit the critical
temperature as a function of the reduced inverse size of the system
for different values of the chemical potential, for $D=3$ and $D=4$
respectively. We see from them that the behavior of the critical
temperature is different, by changing the values of the chemical
potential, for small and large values of $\chi$ (large and small
sizes of the system, respectively). An interesting aspect,
explicitly shown in Figs.~\ref{figPhi41} and \ref{figD4x}, is the
existence of a particular size of the system, $L_0$, corresponding
to the reduced inverse size $\chi_0$, where the critical temperature
vanishes. This particular value $\chi_0$ is {\it independent} of the
chemical potential for both $D=3$ and $D=4$. This is emphasized in
the lower plots of Figs.~\ref{figPhi41} and \ref{figD4x} which show
in detail the domain around the characteristic value
$\chi=\chi_{0}$. For the value of the reduced coupling constant we
take ($\lambda^{\prime} _{R}=0.50$), we find $\chi_0\approx 69.96$
for $D=3$ and  $\chi_0\approx 7.78$ for $D=4$. 

{ Although these results seems {\it a priori} unexpected, we can find  the above
values of $\chi_0$, for both $D=3$ and $D=4$, 
directly from Eq.~(\ref{massa14}). Indeed, it should be noted that for $\chi=\chi_0$, the symmetry 
breaking region disapears completely and we have a null critical temperature. Then $\chi_0$ 
is obtained by solving Eq.~(\ref{massa14}) for $t=0$, in which case only survives the second term, in such a way 
 that all dependency coming from the chemical potential drops out. 
Therefore, for this particular point, $t=0$, for all sizes lower than a characteristic 
value, the chemical potential magnitude does not have any influence. Essentially 
what appears to happen, is that in this case, the finite-size behavior of the physical system 
 collapses into the one corresponding to a zero chemical 
potential, as is the case for a Bose-Einstein transition.} 

We also
see from Figs~\ref{figPhi41} and \ref{figD4x} that, for each value
of $\omega$, there is a limiting smallest size of the system,
$L_{\rm{min}}(\omega)$, corresponding to a largest reduced inverse
size $\chi_{\rm{max}}(\omega)$, such that
$\chi_{\rm{max}}(\omega)>\chi_0$, over which the transition ceases
to exist.

Moreover, we can see clearly from Figs.~\ref{figPhi41} and
\ref{figD4x} that, by effect of the spatial boundaries, $\chi_0$ is
the border between two regions: $\chi <\chi_0$ and $\chi_0
<\chi<\chi_{\rm{max}}$, with different behaviors. In the first
region, the critical temperature is uniquely defined in terms of the
size of the system and of the chemical potential: For each pair
$(\chi,\omega)$ there is only one critical temperature, while in the
second region, two values of $t_c$ may exist for the same values of
$\omega$ and  $\chi$. In the region $\chi_0<\chi<\chi_{\rm{max}}$,
there are for each value of $\omega$, two possible critical
temperatures, say, $t_c^{(1)}$ and $t_c^{(2)}$,  with
$t_c^{(2)}>t_c^{(1)}$, associated respectively to the lower and the
upper branches of the critical curve. This means that in this region
we have two possible transitions. For $D=3$, we take, from
inspection of Fig.~\ref{figPhi41}, $t=t_c^{(1)}=1.93$,
$t=t_c^{(2)}=6.89$ and an intermediate temperature, $t_c^{(1)}<
t<t_c^{(2)}$, $t=4.80$. For these temperatures we plot in
Fig.~\ref{figPhi44}, curves of the effective potential of the
system, given by $U(\beta ,L,\mu ) = \frac{1}{2}m^{2}(\beta ,L,\mu
)\varphi^{2}+\frac{1}{24}\lambda^{\prime} _{R}(\beta ,L,\mu )\varphi
^{4}$ (we define the quantity $\varphi$ by taking $\varphi^2 =
\varphi_a\varphi_a$), for fixed values of the reduced chemical
potential, $\omega=0.30$ and of the reduced inverse size of the
system, $\chi=100$. These plots confirm that two of the
temperatures, $t=t_c^{(1)}=1.93$ (full line) and $t=t_c^{(2)}=6.89$
(dashed line), are critical temperatures corresponding respectively
to the lower and upper branches of the critical curve; the
intermediate temperature, $t_c^{(1)}< t<t_c^{(2)}$, $t=4.80$ (dotted
line), corresponds to the system in the symmetry-broken region. A
similar kind of ``doubling" occurs for large sizes (but {\it not for
$ L\rightarrow \infty $}) of the system. In this case two different
sizes of the system with the same chemical potential may correspond
to the same critical temperature. In the region $\chi<\chi_0$, the
critical temperature grows for increasing chemical potential  for
fixed $\chi$ and also for the system in unlimited space ($
L\rightarrow \infty $).  In this case, we may find the bulk reduced
critical temperature $t_{c}^{\mathrm{bulk}}(\mu ) $ taking the limit
$\chi\rightarrow 0$ in eq.~(\ref{massa14}).

As an overall conclusion, the results suggest that finite-size
effects with finite chemical potential are relevant and deeply
changes the critical curves with respect to the ones for the system
in bulk form. In particular, these actors lead to the appearance of
a ``doubling" of critical parameters, which, up to our knowledge, is
not a trivially expected behavior. This behavior is to be contrasted
with what happens with the system in bulk form, where there is
always an unique critical temperature, which grows with increasing
chemical potential. We also show the existence of a characteristic
size  of the system, which determines the existence, for each value
of the chemical potential, of a minimal size sustaining the broken
phase. {This characteristic size, obtained from Eq.~(\ref{massa14}), 
is the same for all values of the chemical potential.}

\bigskip

\textbf{Acknowledgments}: The authors thank
CAPES, CNPq and FAPERJ, for financial support.

\end{document}